\documentclass[preprint,aps,prd,showpacs,superscriptaddress,nofootinbib,tightenlines]{revtex4-1}

\usepackage{amsfonts}
\usepackage{multirow}
\usepackage{mathrsfs}
\usepackage{graphicx}
\usepackage{amsmath}
\usepackage{amssymb}
\usepackage{bm}
\usepackage{color}
\usepackage{hyperref}
\usepackage{slashed}
\usepackage{float}
\makeatletter
\@ifundefined{showcaptionsetup}{}{%
 \PassOptionsToPackage{caption=false}{subfig}}
\usepackage{subfig}
\makeatother

\usepackage{ulem}

\newcommand{\nn}{\nonumber}

\newcommand{\beq}{\begin{equation}}
\newcommand{\eeq}{\end{equation}}
\newcommand{\bseq}{\begin{subequations}}
\newcommand{\eseq}{\end{subequations}}

\newcommand{\red}[1]{{\color{red}#1}}

\allowdisplaybreaks[4]

\begin{document}

% ------------------------------------------------------------------------------
% Title, authors and institutes
% ------------------------------------------------------------------------------

\title{\mbox{}\\[14pt]
Photoproduction of C-even quarkonia at EIC and EicC
}
%%%%%%%%%%%%%%%%%%%%%%%%%%%%%%%%%%%%%%%%%%%%%%%%%%%%%%%%%%%%%%%%%%%%%%%%%%%%%%

\author{Yu Jia~\footnote{jiay@ihep.ac.cn}}
\affiliation{Institute of High Energy Physics, Chinese Academy of
Sciences, Beijing 100049, China\vspace{0.2 cm}}
\affiliation{School of Physics, University of Chinese Academy of Sciences,
Beijing 100049, China\vspace{0.2 cm}}

\author{Zhewen Mo~\footnote{mozw@ihep.ac.cn}}
\affiliation{Institute of High Energy Physics, Chinese Academy of
	Sciences, Beijing 100049, China\vspace{0.2 cm}}
\affiliation{School of Physics, University of Chinese Academy of Sciences,
	Beijing 100049, China\vspace{0.2 cm}}

\author{Jichen Pan~\footnote{panjichen@ihep.ac.cn}}
\affiliation{Institute of High Energy Physics, Chinese Academy of
	Sciences, Beijing 100049, China\vspace{0.2 cm}}
\affiliation{School of Physics, University of Chinese Academy of Sciences,
	Beijing 100049, China\vspace{0.2 cm}}

\author{Jia-Yue Zhang~\footnote{zhangjiayue@ihep.ac.cn}}
\affiliation{Institute of High Energy Physics, Chinese Academy of
	Sciences, Beijing 100049, China\vspace{0.2 cm}}
\affiliation{School of Physics, University of Chinese Academy of Sciences,
	Beijing 100049, China\vspace{0.2 cm}}

\date{\today}
% ------------------------------------------------------------------------------
% End
% ------------------------------------------------------------------------------

%%%%%%%%%%%%%%%%%%%%%%%%%%%%%%%%%%%%%%%%%%%%%%%%%%%%%%%%%%%%%%%%%%%%%%%%%%%%%%
\begin{abstract}
The $\eta_c$ photoproduction in $ep$ collision has long been proposed as an ideal process to probe the existence of odderon.
In the current work,
we systematically investigate the photoproduction of various $C$-even heavy quarkonia (exemplified by $\eta_{c(b)}$, and $\chi_{c(b)J}$ with $J=0,1,2$) via one-photon exchange channel, at the lowest order in $\alpha_s$ and heavy quark velocity in the context of NRQCD factorization.
We find that the photoproduction rates of the $C$-even quarkonia through this mechanism are comparable in magnitude with that through the
odderon-initiated mechanism, even in the Regge limit ($s\gg -t$), though the latter types of predictions suffers from considerable
theoretical uncertainties.  The future measurements of these types of quarkonium photoproduction processes in \texttt{EIC} and \texttt{EicC} are crucial
to ascertain which mechanism plays the dominant role.
\end{abstract}
%%%%%%%%%%%%%%%%%%%%%%%%%%%%%%%%%%%%%%%%%%%%%%%%%%%%%%%%%%%%%%%%%%%%%%%%%%%%%%
%\pacs{}

\maketitle

\section{Introduction}\label{sec:intro}

Recently, the \texttt{GlueX} experiment at \texttt{JLab} reported the first measurement of near-threshold $J/\psi$ photoproduction off the proton~\cite{Ali:2019lzf},
which triggered a flurry of theoretical investigations. The central issue is whether one can unambiguously infer the QCD trace anomaly contributions to the proton mass
through this ``golden" process. The answer is unclear by far and intensive debates are still going on.  The projected \texttt{SoLID} experiment~\cite{Chen:2014psa} is planned
to accumulate much higher luminosity for the process $\gamma p\to J/\psi+p$ near threshold. Moreover, future experiments at \texttt{EIC}~\cite{AbdulKhalek:2021gbh} and \texttt{EicC}~\cite{Anderle:2021wcy}, designed for the integrated luminosity greater than order $10\;\text{fb}^{-1}$, can measure both
$J/\psi$ and $\Upsilon$ photo- and lepto-production near threshold with better accuracy.

Besides the vector quarkonium near-threshold photoproduction, another type of exclusive quarkonium production processes in $ep$ collision is also of great theoretical
interest. Concretely speaking, it has long been proposed that the $C$-even quarkonium such as $\eta_c$ photoproduction in the Regge limit ({\it i.e.},
large $s$ yet small $t$ limit) is an ideal place to look for the existence of the odderon.

As is well known, long before the advent of QCD, elastic hadron scattering processes in the forward limit have been
analysed in the framework of Regge theory~\cite{Ewerz:2003xi}, which is built upon some general axiomatic principles of
$S$-matrix theory such as analyticity and unitarity.
Two central objects in Regge theory are the Reggeized multi-gluon compounds responsible for $t$-channel exchange, with the $C$-even state called pomeron~\cite{Sibirtsev:2004ca,Donnachie:1988nj,Donnachie:1994zb,Laget:1994ba,Donnachie:1999qv}, while the $C$-odd one called odderon~\cite{Lukaszuk:1973nt}.
In the present work, we concentrate on the effect of odderon. The $C$-odd property of odderon indicates there is a difference between the $pp$ and $p \bar{p}$ elastic
scattering in the Regge limit. Apart from early experimental evidence~\cite{Breakstone:1985pe,Erhan:1984mv,Avila:2006wy},
a recent careful analysis has been conducted to compare the $p \bar{p}$ elastic cross section measured in \texttt{D0} experiment~\cite{D0:2012erd} at $\sqrt{s}=1.96\:\text{TeV}$,
with the $pp$ elastic cross section measured at \texttt{LHC} experiment~\cite{Antchev:2017yns} extrapolated to $\sqrt{s}=1.96\:\text{TeV}$, the $3.4\sigma$ level
discrepancy~\cite{Abazov:2020rus} is viewed as strong evidence for the existence of the odderon.

On theoretical ground, Bartels-Kwiecinski-Praszalowicz (BKP) equation~\cite{Kwiecinski:1980wb}, an integral equation for the Green functions of three
Reggeized gluons in the $t$-channel, plays a central role in the odderon theory. Two different solutions to the BKP equation are
Bartels-Lipatov-Vacca(BLV) odderon solution~\cite{Bartels:1999yt} and Janik-Wosiek(JW) odderon solution~\cite{Janik:1998xj,Braun:1998mg},
which differ in the quantum number $q_3$~\cite{Gauron:1993dc,Braun:1998mg} as well as the intercept $\alpha(0)$, characterizing the
scaling behavior of the total cross section with $s$ in the Regge limit, $\sigma \sim s^{\alpha(0)-1}$ arising from Regge pole~\cite{Ewerz:2003xi}.
It appears necessary to examine the different solutions  channel by channel.

\begin{figure}[h]
  \includegraphics{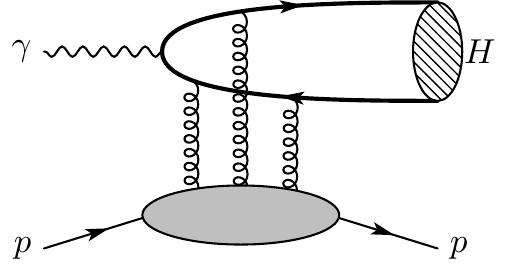}
  \caption{Three gluon/odderon $t$-channel exchange for $\gamma p\to H+p$, where $H$ represents a $C$-even quarkonium.}
  \label{fig:diagrams_gluon}
\end{figure}

According to \cite{Bartels:2001hw,Ewerz:2004rc}, both BLV and JW solutions contribute to the $p p$/$p \bar{p}$ scattering.
Consequently, concerning only the $p p$/$p \bar{p}$ forward scattering is not sufficient to single out the correct odderon model.
In contrast, for the $C$-even neutral meson photoproduction process, the BLV solution yield a nonvanishing contribution, but the JW solution
does not contribute because this solution vanishes if two of the three gluons are located at same spacetime point. Pseudoscalar/tensor neutral meson photo(electro-)
production process, such as $f_{2}$(1270)~\cite{Berger:2000wt}, $\pi^{0}$~\cite{Adloff:2002dw,Harland-Lang:2018ytk}, $\eta$(548)~\cite{Harland-Lang:2018ytk},
{\it etc.}, is necessary to confirm the BLV solution~\cite{Ewerz:2004rc}. Among them, the pseudoscalar charmonium $\eta_{c}$
photoproduction  has attracted much attention~\cite{Czyzewski:1996bv,Bartels:2001hw,Bartels:2003zu,Vacca:2001ir,Engel:1997cga,Goncalves:2015hra,Bartels:1999yt,Kilian:1997ew,Berger:1999rq}.
A representative diagram for the odderon contribution to $\gamma p\to H p$ ($H$ representing a $C$-even quarkonium) is shown in FIG.\ref{fig:diagrams_gluon}.

It is fair to note that the aforementioned odderon-based models are subjected to large theoretical uncertainties.
Needless to say, it is crucial for the future \texttt{EIC} and \texttt{EicC} experiments to provide the key examination.
On the other hand, another important mechanism for the $C$-even quarkonium photoproduction, {\it i.e.}, through one-photon $t$-channel exchange,
has not been adequately studied in literature. At least, this one-photon exchange contribution constitutes the important background to
pin down the odderon contribution unambiguously. Moreover, $\gamma p\to H p$ itself is also of theoretical interest,
which provides a novel means to test the applicability of NRQCD effective theory~\cite{Bodwin:1994jh}  for exclusive quarkonium production in $ep$ collision.

It then becomes the central aim of this work to study the photoproduction of $C$-even quarkonium $\gamma p\to H p$ through one-photon exchange,
in the framework of NRQCD factorization. The calculation is conducted at lowest order in $\alpha_s$ and heavy quark velocity expansion.
We will consider $H$ to be both $S$-wave spin-singlet quarkonia $\eta_{c,b}$ and $P$-wave spin-triplet quarkonia  $\chi_{cJ}$ ($\chi_{bJ}$) ($J$=0,1,2).
Our main finding is that this type of contribution may be comparable in magnitude with the odderon contribution in the Regge limit, thus
can not be neglected. We hope the future \texttt{EIC} and \texttt{EicC} experiments can measure these processes and explicitly test our predictions.

The structure of this paper is distributed as follows.
%-------------------------------
In Section~\ref{sec:photon}, we present the formulae of differential cross sections for the
$\gamma p\to H p$ processes through one-photon exchange mechanism, with
$H=\eta_{c,b}$, $\chi_{cJ}$ ($\chi_{bJ}$) ($J$=0,1,2). The predictions are given by the lowest order calculation
within the NRQCD factorization framework.
%-------------------------------
In Section~\ref{sec:asymptotic}, we discuss the asymptotic behaviors of the differential cross sections
in Regge limit and near-threshold limit, respectively.
%-------------------------------
Section~\ref{sec:numerical} is devoted to the numerical results and phenomenological analyses.
%-------------------------------
We conclude in Section~\ref{sec:conclusion}.
For the sake of completeness, yet mainly for amusement, in Appendix~\ref{sec:graviton} we also present the explicit expressions of
the differential cross section for $\gamma p\to J/\psi+p$ from the one-graviton $t$-channel exchange (analogous to one-photon exchange).

\section{One-Photon exchange mechanism for
quarkonium photoproduction}\label{sec:photon}

\begin{figure}[h]
  \includegraphics{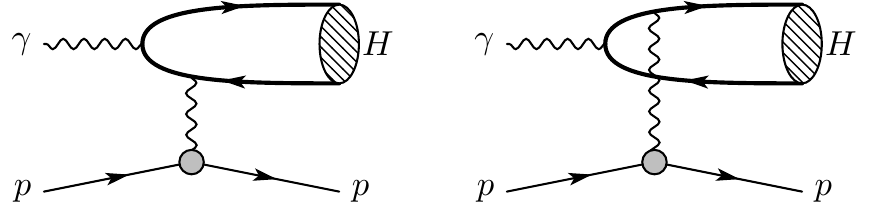}
  \caption{Two lowest-order Feynman diagrams for $C$-even quarkonium photoproduction. }
  \label{fig:diagrams_photon}
\end{figure}

Let us turn to the photoproduction process $\gamma p\rightarrow Hp$, with $H$ signalling the $C$-even quarkonium state $H$
exemplifying $\eta_{c}$ and $\chi_{cJ}$. The same analysis also applies to
bottomonia photoproduction. The leading-order (LO) Feynman diagrams that incorporates the one-photon exchange are
depicted in FIG. \ref{fig:diagrams_photon}.  
The amplitude possesses the following factorized structure:
%-------------------------------
\beq
%-------------------------------
\mathcal{M}=\frac{e^2 g_{\mu\nu}}{t}\langle H\left(P\right)\vert J_\text{EM}^{\mu}\vert\gamma\left(k\right)\rangle\langle p\left(P_2\right)\vert J^\nu_\text{EM}\vert p\left(P_1\right)\rangle,
\label{eq:iM}
%-------------------------------
\eeq
%-------------------------------
where the external momenta of each particle are specified in the parentheses, the momentum 
carried by the virtual photon is $q=P-k=P_1-P_2$, and $t\equiv q^{2}\equiv-Q^2$. 
The electromagnetic current $J^\mu_\text{EM}$ bears the standard definition: 
%-------------------------------
\beq
%-------------------------------
  J^\mu_\text{EM} = \sum_f e_f\bar{q}_f\gamma^\mu q_f+\cdots,
%-------------------------------  
\eeq
%-------------------------------
where we have only retained the quark sector contribution, $e_{u}=2/3$ for up-type quark , and $e_d=-1/3$ for down-type quark.

\begin{figure}
  \includegraphics[scale=0.6]{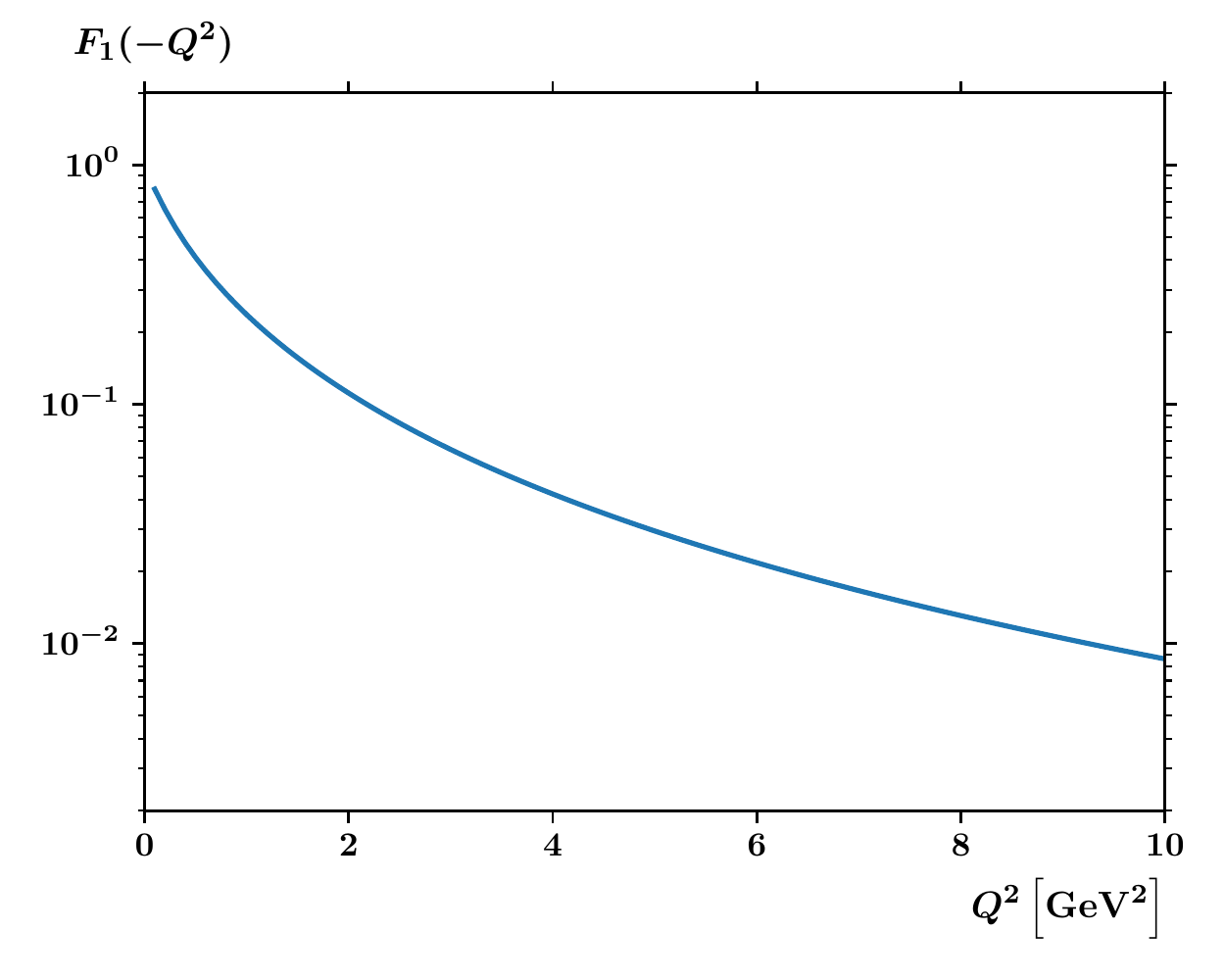}
  \includegraphics[scale=0.6]{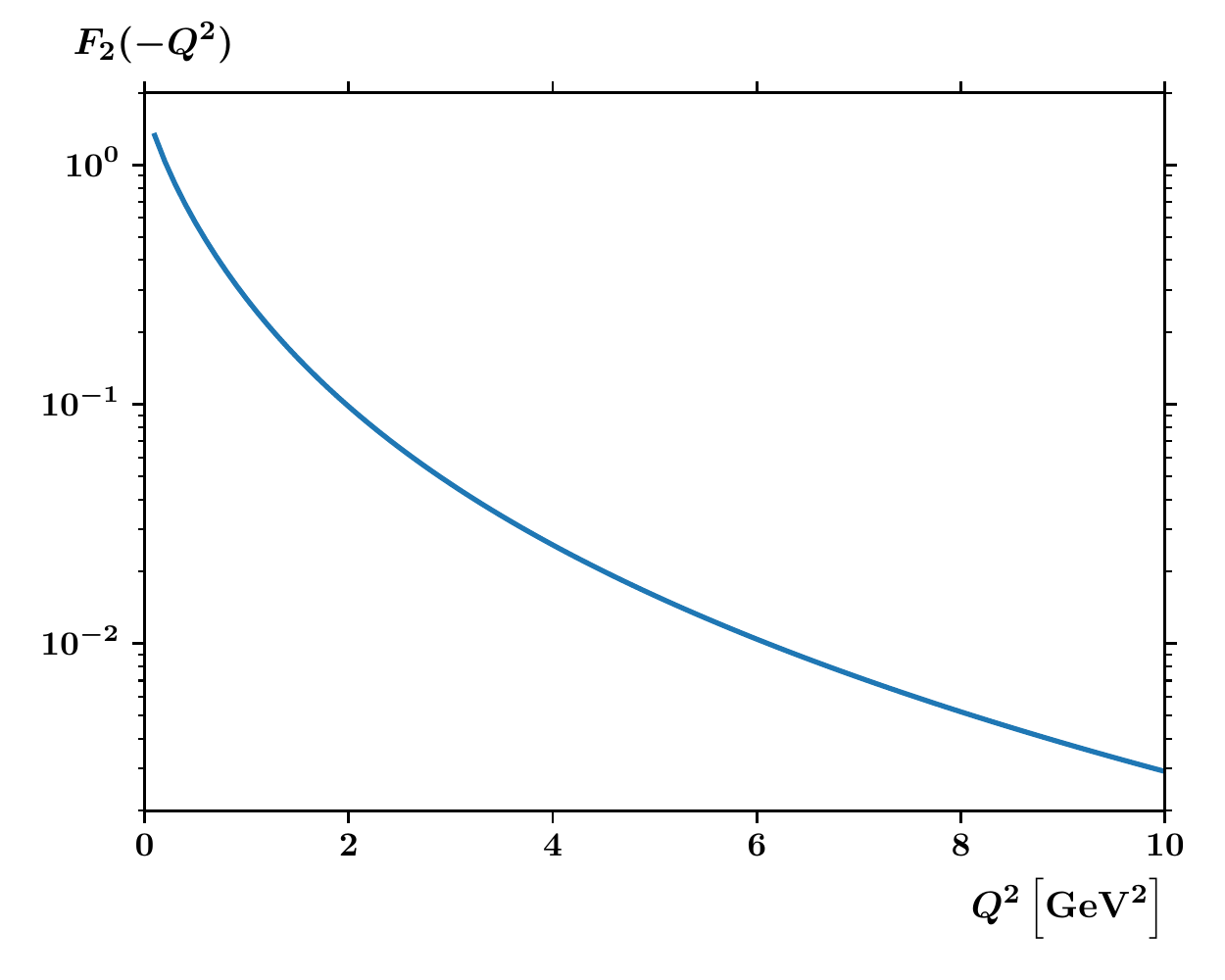}
  \caption{ The experimentally-determined proton electromagnetic form factors $F_1$ and $F_2$ 
  as a function of $Q^2$~\cite{Ye:2017gyb}. }
  \label{fig:form_factor}
\end{figure}

The electromagnetic vertex in the lower half of each Feynman diagram in FIG. \ref{fig:diagrams_photon} encapsulates the contribution from the 
familiar proton electromagnetic form factors:
%-------------------------------
\beq
  \langle p\left(P_2\right)\vert J_\text{EM}^{\mu}\vert p\left(P_1\right)\rangle=\bar{u}\left(P_2\right)\left[\gamma^{\mu}F_{1}\left(q^{2}\right)+\frac{i\sigma^{\mu\nu}q_{\nu}}
  {2M_{p}}F_{2}\left(q^{2}\right)\right]u\left(P_1\right),
  \label{eq:p_current}
\eeq
%-------------------------------
where the form factors $F_1\left(q^2\right)$ and $F_2\left(q^2\right)$ are real-valued functions.
The perturbative QCD analysis reveals that the form factors scale asymptotically as
$F_1(q^2) \sim 1/q^4$~\cite{Brodsky:1973kr, Lepage:1979za}, and $F_2(q^2)/F_1(q^2) \sim \log^2 (q^2/\Lambda^2)/q^2$~\cite{Belitsky:2002kj}. 
An explicit plot of the proton form factors versus momentum transfer, obtained through fitting a large set of data~\cite{Ye:2017gyb},  
is displayed in FIG. \ref{fig:form_factor},
which we will use for later numerical analysis.

The $\gamma \gamma^* \to H$ vertex in the upper half of each diagram in FIG. \ref{fig:diagrams_photon} encodes the photon-to-charmonium electromagnetic transition form factor.
Irrespective of the explicit value of $Q^2$, owing to the fact $m_c\gg \Lambda_{\rm QCD}$ and asymptotic freedom, 
these EM transition form factors can be factorized in the product of the short-distance coefficients and long-distance NRQCD matrix elements.
At the lowest order in $\alpha_s$ and
$v$, a straightforward calculation leads to the following expressions for various photon-to-$H$ electromagnetic transition form factors, 
with $H=\eta_{c},\chi_{c0,1,2}$:
%-------------------------------
\bseq
%-------------------------------
\begin{align}
  & \langle\eta_{c}\left(P\right)\vert J_\text{EM}^{\mu}\vert\gamma\left(k\right)\rangle=-\frac{4ie_{c}e}{m_{c}^{1/2}\left(4m_{c}^{2}-t\right)}\sqrt{\frac{N_{c}}{2\pi}}R_{S}\left(0\right)\epsilon^{\mu\nu\rho\sigma}\varepsilon_{\nu}\left(k\right)k_{\rho}P_{\sigma},\\
  & \langle\chi_{c0}\left(P\right)\vert J_\text{EM}^{\mu}\vert\gamma\left(k\right)\rangle=\frac{2\sqrt{3}e_{c}e}{3m_{c}^{3/2}\left(4m_{c}^{2}-t\right)^{2}}\sqrt{\frac{3N_{c}}{2\pi}}R_{P}^{\prime}\left(0\right)\left(12m_{c}^{2}-t\right)\\\nn
  &\qquad\times\left[\left(4m_{c}^{2}-t\right)g^{\mu\nu}-2k^{\mu}P^{\nu}\right]\varepsilon_{\nu}\left(k\right),\\
  & \langle\chi_{c1}\left(P\right)\vert J_\text{EM}^{\mu}\vert\gamma\left(k\right)\rangle=-\frac{\sqrt{2}ie_{c}e}{m_{c}^{5/2}\left(4m_{c}^{2}-t\right)^{2}}\sqrt{\frac{3N_{c}}{2\pi}}R_{P}^{\prime}\left(0\right)\varepsilon_{\alpha}^{*}\left(P\right)\Big\{2\left[4m_{c}^{2}k^{\mu}+\left(t-4m_{c}^{2}\right)P^{\mu}\right]\nonumber \\
  & \qquad\times\epsilon^{\nu\rho\sigma\alpha}k_{\rho}P_{\sigma}-2tP^{\nu}\epsilon^{\mu\rho\sigma\alpha}k_{\rho}P_{\sigma}+t\left(4m_{c}^{2}-t\right)\epsilon^{\mu\nu\sigma\alpha}P_{\sigma}\Big\}\varepsilon_{\nu}\left(k\right),\\
  & \langle\chi_{c2}\left(P\right)\vert J_\text{EM}^{\mu}\vert\gamma\left(k\right)\rangle=\frac{e_{c}e}{3m_{c}^{7/2}\left(4m_{c}^{2}-t\right)^{2}}\sqrt{\frac{3N_{c}}{2\pi}}R_{P}^{\prime}\left(0\right)\varepsilon_{\alpha\beta}^{*}\left(P\right)\Big\{2m_{c}^{2}\left(12m_{c}^{2}-t\right)\nn\\
  &\qquad\times\big[\left(4m_{c}^{2}-t\right)g^{\mu\nu}-2k^{\mu}P^{\nu}\big]g^{\alpha\beta}+96m_{c}^{4}g^{\mu\nu}k^{\alpha}k^{\beta}-t\left[\left(4m_{c}^{2}-t\right)g^{\mu\nu}-2k^{\mu}P^{\nu}\right]P^{\alpha}P^{\beta}\nonumber \\
  & \qquad-12m_{c}^{2}\left[\left(4m_{c}^{2}-t\right)g^{\mu\nu}+\left(P^{\mu}-k^{\mu}\right)P^{\nu}\right]\left(k^{\alpha}P^{\beta}+k^{\beta}P^{\alpha}\right)\nonumber \\
  & \qquad+48m_{c}^{4}P^{\nu}\left(k^{\alpha}g^{\beta\mu}+k^{\beta}g^{\alpha\mu}\right)-48m_{c}^{4}k^{\mu}\left(k^{\alpha}g^{\beta\nu}+k^{\beta}g^{\alpha\nu}\right)\nonumber \\
  & \qquad+6m_{c}^{2}\left(4m_{c}^{2}-t\right)\left(k^{\mu}+P^{\mu}\right)\left(P^{\alpha}g^{\beta\nu}+P^{\beta}g^{\alpha\nu}\right)\nonumber \\
  & \qquad-24m_{c}^{4}\left(4m_{c}^{2}-t\right)\left(g^{\alpha\mu}g^{\beta\nu}+g^{\alpha\nu}g^{\beta\mu}\right)\Big\}\varepsilon_{\nu}\left(k\right),
\end{align}
%-------------------------------
\label{eq:H_current}
%-------------------------------
\eseq
%-------------------------------
where $N_c=3$ is the number of colors, and $\varepsilon_{\alpha}(P),\varepsilon_{\alpha\beta}(P)$ represent the polarization vector and tensor of
$\chi_{c1},\,\chi_{c2}$, respectively. One readily verifies that the expressions in \eqref{eq:H_current}
obey the conservation of the current $J_\text{EM}^\mu$, Ward identity for the incoming photon, as well as the discrete
$C,P,T$ symmetries. In passing, we note that the $\gamma$-to-$\eta_c$ transition form factor has already been computed to 
${\cal O}(\alpha_s^2)$~\cite{Feng:2015uha}.
For later phenomenological analysis, it is convenient to approximate the nonperturbative vaccum-to-$H$ NRQCD 
matrix elements by the (first derivative of) 
radial wave functions at the origin $R_S(0)$ ($R^\prime_P(0)$) in potential model.

Starting from the standard formula~\cite{pdg:2020}:
%-------------------------------
\beq
  {d\sigma\left(\gamma p\rightarrow H+p\right)\over dt}=\frac{1}{64\pi s}\frac{1}{\left|\mathbf{k}_{\text{cm}}\right|^{2}}\frac{1}{2\times2}\sum_{\text{Polar.}}\left|\mathcal{M}\right|^{2},
\eeq
%-------------------------------
(with $\mathbf{k}_{\text{cm}}$ referring to the magnitude of the photon momentum in the center-of-mass frame), 
combined with \eqref{eq:p_current} and \eqref{eq:H_current},
we then obtain the desired differential rates for photoproduction of $H$:
%-------------------------------
\bseq
%-------------------------------
\begin{align}
  & \frac{d\sigma\left(\gamma p\rightarrow\eta_{c}p\right)}{dt}=
  \frac{4\pi e_{c}^{4}\alpha^{3}N_{c}\left|R_{S}\left(0\right)\right|^{2}}{m_{c}t^{2}\left(t-4m_{c}^{2}\right)^{2}\left(M_{p}^{2}-s\right)^{2}}
  \Big\{2\left[8m_{c}^{2}t\left(M_{p}^{2}+s+t\right)-16m_{c}^{4}\left(2M_{p}^{2}+t\right)\right. \nonumber \\
  & \qquad-t\left(2M_{p}^{4}-4M_{p}^{2}s+2s^{2}+2st+t^{2}\right)]F_{1}^2(t)-4t\left(t-4m_{c}^{2}\right)^{2}F_{1}(t)F_{2}(t)\nonumber \\
  & \qquad-t\left[16m_{c}^{4}+4m_{c}^{2}t\left(\dfrac{s}{M_{p}^{2}}-3\right)+t\left(-M_{p}^{2}+2\left(s+t\right)-\dfrac{s\left(s+t\right)}{M_{p}^{2}}\right)\right]F_{2}^{2}(t)\Big\},\\
  & \frac{d\sigma\left(\gamma p\rightarrow\chi_{c0}p\right)}{dt}=\frac{\left|R_{P}^{\prime}\left(0\right)\right|^{2}}{m_{c}^{2}\left|R_{S}\left(0\right)\right|^{2}}\left(\frac{12m_{c}^{2}-t}{4m_{c}^{2}-t}\right)^{2}\frac{d\sigma\left(\gamma p\rightarrow\eta_{c}p\right)}{dt},\\
  & \frac{d\sigma\left(\gamma p\rightarrow\chi_{c1}p\right)}{dt}=\frac{24\pi e_{c}^{4}\alpha^{3}N_{c}\left|R_{P}^{\prime}\left(0\right)\right|^{2}}{m_{c}^{3}\left(t-4m_{c}^{2}\right)^{4}\left(M_{p}^{2}-s\right)^{2}}
  \Big\{\Big[-t\left(-4sM_{p}^{2}+2M_{p}^{4}+2s^{2}+2st+t^{2}\right)\nonumber \\
  & \qquad+8m_{c}^{2}\left(-M_{p}^{2}(4s+t)+2M_{p}^{4}+2s^{2}+3st+t^{2}\right)-16m_{c}^{4}\left(-2M_{p}^{2}+4s+t\right)\Big]F_{1}^2(t)\nonumber \\
  & \qquad -4\left(t-4m_{c}^{2}\right)^{2}\left(t+4m_{c}^{2}\right)F_{1}(t)F_{2}(t)\nonumber \\
  & \qquad+\Big[-2tm_{c}^{2}\left(-2(4s+3t)+4M_{p}^{2}+\dfrac{4s^{2}+6st+t^{2}}{M_{p}^{2}}\right)+16tm_{c}^{4}\left(1+\dfrac{2s+t}{M_{p}^{2}}\right)\nonumber \\
  & \qquad-32m_{c}^{6}\left(4+\dfrac{t}{M_{p}^{2}}\right)-t^{2}\left(2(s+t)-M_{p}^{2}-\dfrac{s(s+t)}{M_{p}^{2}}\right)\Big]F_{2}^{2}(t)\Big\},\\
  & \frac{d\sigma\left(\gamma p\rightarrow\chi_{c2}p\right)}{dt}=\frac{8\pi e_{c}^{4}\alpha^{3}N_{c}\left|R_{P}^{\prime}\left(0\right)\right|^{2}}{m_{c}^{3}t^{2}\left(t-4m_{c}^{2}\right)^{4}\left(M_{p}^{2}-s\right)^{2}}
  \Big\{2\Big[8t^{2}m_{c}^{2}\left(-M_{p}^{2}(12s+5t)\right.\nonumber \\
  & \qquad \left.+6M_{p}^{4}+6s^{2}+7st+t^{2}\right)-16tm_{c}^{4}\left(-2M_{p}^{2}(12s+5t)+12M_{p}^{4}+12s^{2}+24st+7t^{2}\right)\nonumber \\
  & \qquad+768tm_{c}^{6}\left(M_{p}^{2}+s+t\right)-1536m_{c}^{8}\left(2M_{p}^{2}+t\right)+t^{3}\left(-4sM_{p}^{2}+2M_{p}^{4}+2s^{2}+2st+t^{2}\right)\Big]\nonumber \\
  & \qquad F_{1}^2(t)-4t\left(t-4m_{c}^{2}\right)^{2}\left(12tm_{c}^{2}+96m_{c}^{4}+t^{2}\right)F_{1}(t)F_{2}(t)\nonumber \\
  & \qquad-t\Big[-16tm_{c}^{4}\left(-(12s+7t)+6M_{p}^{2}+\dfrac{3\left(2s^{2}+4st+t^{2}\right)}{M_{p}^{2}}\right)\nonumber \\
  & \qquad+2t^{2}m_{c}^{2}\left(-6(4s+t)+12M_{p}^{2}+\dfrac{12s^{2}+14st+3t^{2}}{M_{p}^{2}}\right)-96tm_{c}^{6}\left(8-\dfrac{4s+t}{M_{p}^{2}}\right)\nonumber \\
  & \qquad+1536m_{c}^{8}+t^{3}\left(2(s+t)-M_{p}^{2}-\dfrac{s(s+t)}{M_{p}^{2}}\right)\Big]F_{2}^{2}(t)\Big\}.
%-------------------------------
\end{align}
%-------------------------------
\label{eq:sigma}
%-------------------------------
\eseq
%-------------------------------

The kinematically allowed region for this $2$-to-$2$ process at a given $\sqrt{s}$ 
is \red{$t_\text{min}\le t \le  t_\text{max}$}, with
%-------------------------------
\bseq
%-------------------------------
\begin{align}
%-------------------------------
  t_{\text{min}} & =\frac{M_H^{2}\left(s+M_{p}^{2}\right)-\left(s-M_{p}^{2}\right)\left(s-M_{p}^{2}-\sqrt{\lambda\left(s,M_{p}^{2},M_H^{2}\right)}\right)}{2s},\\
  t_{\text{max}} & =\frac{M_H^{2}\left(s+M_{p}^{2}\right)-\left(s-M_{p}^{2}\right)\left(s-M_{p}^{2}+\sqrt{\lambda\left(s,M_{p}^{2},M_H^{2}\right)}\right)}{2s}.
\end{align}
\label{tmin:and:tmax}
\eseq
%-------------------------------
$M_H$ denotes the mass of hadron, and $\lambda(x,y,z)=x^2+y^2+z^2-2xy-2yz-2zx$ is the K\"allen function. 
One may see \cite{Guo:2021ibg} for a more detailed discussion.

\section{Asymptotic behavior}\label{sec:asymptotic}

\subsection{Regge limit}

In this work we are primarily interested in the Regge limit.  
If we presume this limit is set by $s \gg (2m_c)^2 \gg |t|$,   
one then anticipates that the differential rate near $t_\mathrm{min} \to -M_p^2(2m_c)^4/s^2$ 
(obtained from the explicit definition of (\ref{tmin:and:tmax})) is the desired asymptotic behavior. 
However, by closely inspecting the actual numerical results shown in FIG.~\ref{fig:cross_c_photon_W50} and FIG.~\ref{fig:cross_b_photon_W50}, 
one sees that the differential production rates fall off rather stiffly as $t \to t_\mathrm{min}$. 
Therefore, $d\sigma/dt$ at $t=t_\mathrm{min}$ is not representative for the small $|t|$ behavior that one
is really interested in. The asymptotic expressions below are actually derived in the region where
$s \gg (2m_c)^2$ and $|t_\mathrm{min}| \ll |t| \ll m_p^2 < (2m_c)^2$~\footnote{We found that our result of $d\sigma\left(\gamma p\rightarrow\eta_{c}p\right)/dt$ 
differs from \cite{Ma:2003py} in the $s \gg (2m_c)^2$ and small $|t|$ limit.}:
%-------------------------------
\bseq
%-------------------------------
\begin{align}
%-------------------------------
   \frac{d\sigma\left(\gamma p\rightarrow\eta_{c}p\right)}{dt}\approx & \frac{\pi e_{c}^{4}\alpha^{3}N_{c}\left|R_{S}\left(0\right)\right|^{2}F_{1}^{2}\left(0\right)}{m_{c}^{5}\left(-t\right)},\\
   \frac{d\sigma\left(\gamma p\rightarrow\chi_{c0}p\right)}{dt}\approx & \frac{9\pi e_{c}^{4}\alpha^{3}N_{c}\left|R_{P}^{\prime}\left(0\right)\right|^{2}F_{1}^{2}\left(0\right)}{m_{c}^{7}\left(-t\right)},\\
 \frac{d\sigma\left(\gamma p\rightarrow\chi_{c1}p\right)}{dt}\approx & \frac{3\pi e_{c}^{4}\alpha^{3}N_{c}\left|R_{P}^{\prime}\left(0\right)\right|^{2}F_{1}^{2}\left(0\right)}{m_{c}^{9}},\\
  \frac{d\sigma\left(\gamma p\rightarrow\chi_{c2}p\right)}{dt}\approx & \frac{12\pi e_{c}^{4}\alpha^{3}N_{c}\left|R_{P}^{\prime}\left(0\right)\right|^{2}F_{1}^{2}\left(0\right)}{m_{c}^{7}\left(-t\right)}.
%-------------------------------
\end{align}
%-------------------------------
\label{eq:dsigma_asym}
%-------------------------------
\eseq
%-------------------------------

From \eqref{eq:dsigma_asym} one immediately observes that the differential photoproduction rates are
independent of $s$ in the Regge limit.
An interesting pattern is that, in the small $|t|$ limit, the $\eta_c$, $\chi_{c0}$ and $\chi_{c2}$ photoproduction rates scale as $\mathcal{O}(1/t)$,
whereas the $\chi_{c1}$ differential rate exhibits quite differential scaling $\propto t^0$. 
These scaling behaviors can also be clearly visualized in FIG. \ref{fig:cross_c_photon_W50} and FIG. \ref{fig:cross_b_photon_W50},
which can be ascribed to the Landau-Yang theorem.
It should also be emphasized that,
though \eqref{eq:dsigma_asym} are insensitive to $s$ when $\sqrt{s}$ increases, the lower limit of $|t|$ where
\eqref{eq:dsigma_asym} remains applicable continues to decrease toward $t_\mathrm{min}$.
As a result, for the total cross section of $\eta_{c(b)}$, $\chi_{c(b)0}$ and $\chi_{c(b)2}$,
the $t$-integrals of \eqref{eq:dsigma_asym} yield a $\log s$ enhancement, while the total cross
sections for $\chi_{c(b)1}$ photoproduction remains constant in the large $s$ limit, which can be seen from
FIG. \ref{fig:cross_c_photon_total} and FIG. \ref{fig:cross_b_photon_total}.

\subsection{Threshold limit}

It is also interesting to consider another very different limit, $\sqrt{s}\approx M_H+M_p$, 
where the center-of-mass energy is just enough to produce the static $H$ and $p$ in the final 
state. In such as limit, $t$ can be fixed to be $t_0=-4 M_p m_c^2/(M_p+2 m_c)$. 
In the near-thereshold limit, various differential photoproduction rates can be approximated as
%-------------------------------
\bseq
%-------------------------------
\begin{align}
%-------------------------------
   \frac{d\sigma\left(\gamma p\rightarrow\eta_{c}p\right)}{dt}\approx & \frac{\pi e_c^4 \alpha^3 N_c\left|R_{S}\left(0\right)\right|^{2}(F_1(t_0)+F_2(t_0))^2(M_p+2 m_c)}{8 M_p m_c^5 (M_p+m_c)^2},\\
  \frac{d\sigma\left(\gamma p\rightarrow\chi_{c0}p\right)}{dt}\approx & \frac{\pi e_c^4 \alpha^3N_c \left|R_{P}^{\prime}\left(0\right)\right|^{2}(F_1(t_0)+F_2(t_0))^2(M_p+2 m_c)(2 M_p+3 m_c)^2}{8 M_p m_c^7 (M_p+m_c)^4}\\
  \frac{d\sigma\left(\gamma p\rightarrow\chi_{c1}p\right)}{dt}\approx & \frac{3 \pi  e_c^4 \alpha ^3N_c \left|R_{P}^{\prime}\left(0\right)\right|^{2} (M_p+2 m_c)}{16 M_p m_c^9 (M_p+m_c)^4} \left
  (F_1^2(t_0) M_p^2 \left(M_p^2+4 M_p m_c+5 m_c^2\right)\right.\nonumber\\
  & \left.-4 F_1(t_0) F_2(t_0) M_p
   m_c^3+F_2^2(t_0) m_c^2 \left(M_p^2+m_c^2\right)\right),\\
   \frac{d\sigma\left(\gamma p\rightarrow\chi_{c2}p\right)}{dt}\approx &\frac{\pi e_c^4 \alpha ^3 N_c  \left|R_{P}^{\prime}\left(0\right)\right|^{2} (M_p+2 m_c)}{16 M_p m_c^9 (M_p+m_c)^4} \nonumber \\
  & \times\left( F_1^2(t_0) \left(3 M_p^4+12 M_p^3 m_c+19 M_p^2 m_c^2+24 M_p m_c^3+24
   m_c^4\right)\right. \nonumber\\
   & +4 F_1(t_0)F_2(t_0) m_c^2 \left(2 M_p^2+9 M_p m_c+12 m_c^2\right)\nonumber \\
   & \left.+F_2^2(t_0) m_c^2 \left(7 M_p^2+24 M_p
   m_c+27 m_c^2\right)\right).
%-------------------------------
\end{align}
%-------------------------------
\label{eq:dsigma_asym_threshold}
%-------------------------------
\eseq
%-------------------------------
As it should be, the angular distributions of the $H$ are always isotropic in the near-threshold limit,  
and the integrated cross sections are suppressed by the relative velocity between the $H$ and proton $P$, $\propto \sqrt{s-(M_p+M_H)^2}$.

\section{Numerical analysis}\label{sec:numerical}

\subsection{Photoproduction rates}

To make explicit phenomenological predictions,  we first need to fix the values of various input parameters.  
We choose the values of the charmonium (bottomonium) wave function at origin from the Cornell potential model~\cite{Eichten:2019hbb}.
%-------------------------------
\bseq
\begin{align}
  &M_p=0.938\:\text{GeV},\quad m_c=1.5\:\text{GeV},\quad m_b=4.7\:\text{GeV}.\\
  &\left|R_{1S(c\bar{c})}\right|^2=1.0952\:\text{GeV}^3, \quad \left|R_{1P(c\bar{c})}^{\prime}(0)\right|^2=0.1296\:\text{GeV}^5, \\
  &\left|R_{1S(b\bar{b})}\right|^2=5.8588\:\text{GeV}^3, \quad \left|R_{1P(b\bar{b})}^{\prime}(0)\right|^2=1.6057\:\text{GeV}^5.
\end{align}
%-------------------------------
\eseq
%-------------------------------

\begin{figure}[b]

  \subfloat[$\sqrt{s}=4.3\:\text{GeV}$]{\includegraphics[scale=0.6]{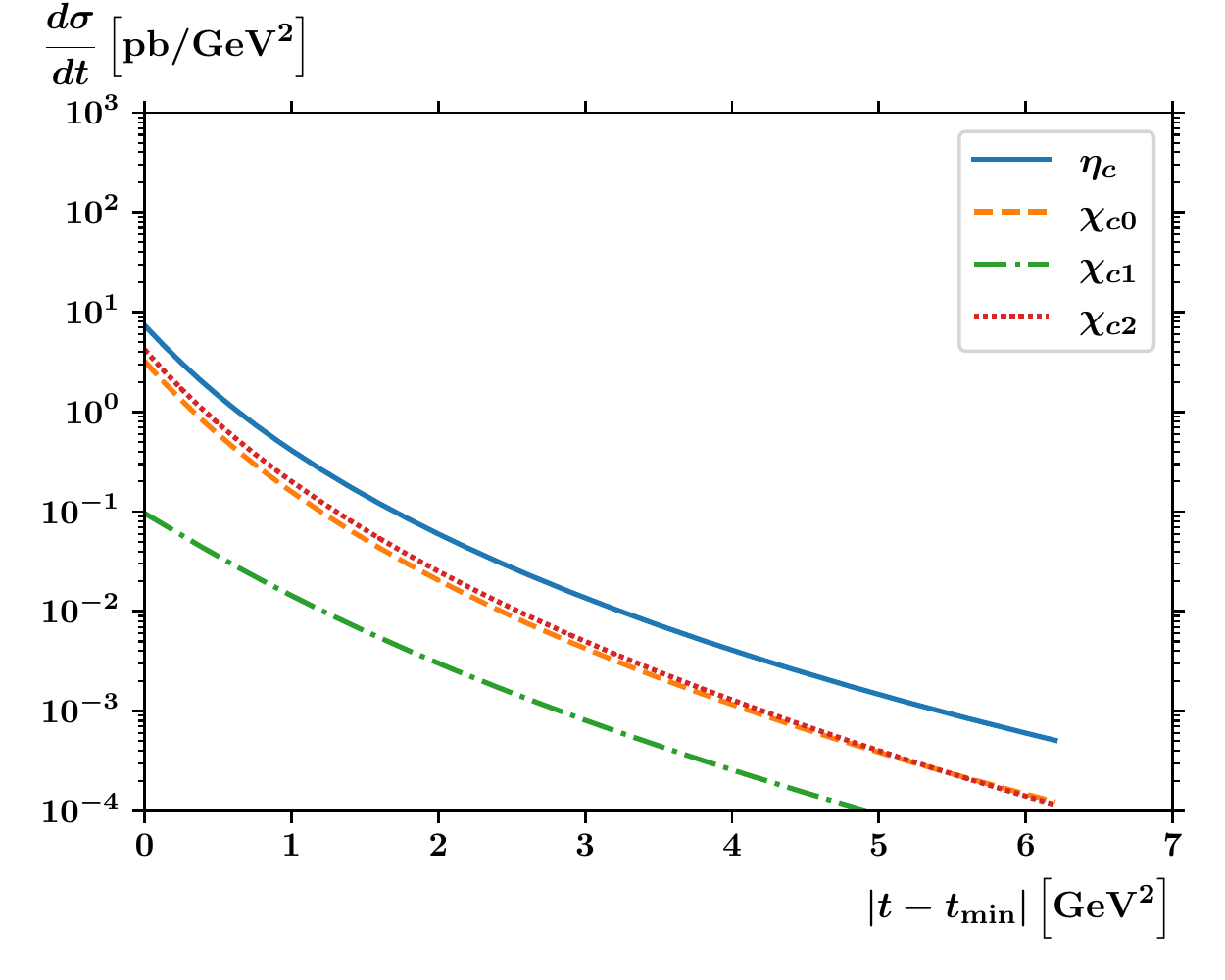}}
  \subfloat[$\sqrt{s}=15\:\text{GeV}$]{\includegraphics[scale=0.6]{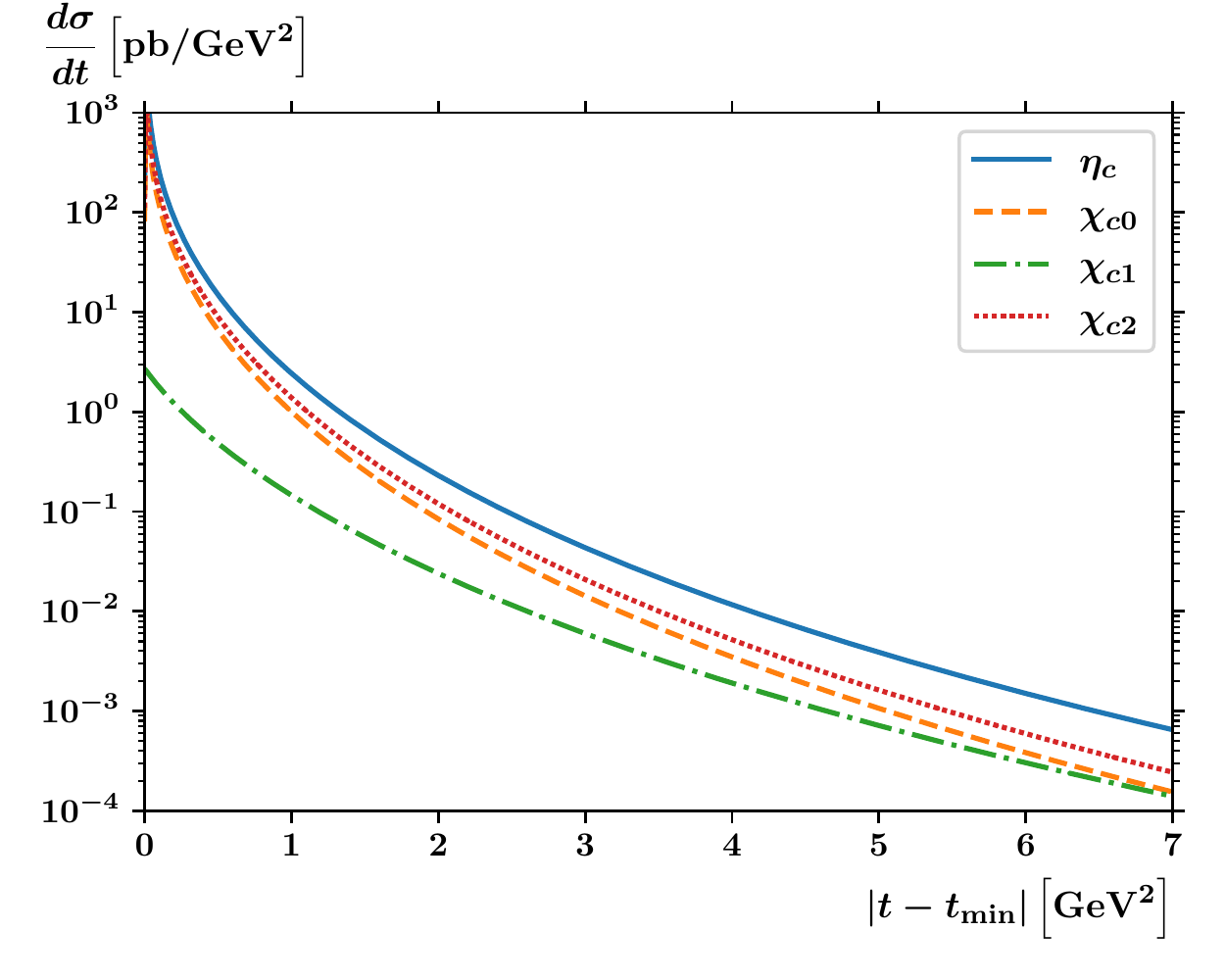}}

  \subfloat[$\sqrt{s}=50\:\text{GeV}$]{\includegraphics[scale=0.6]{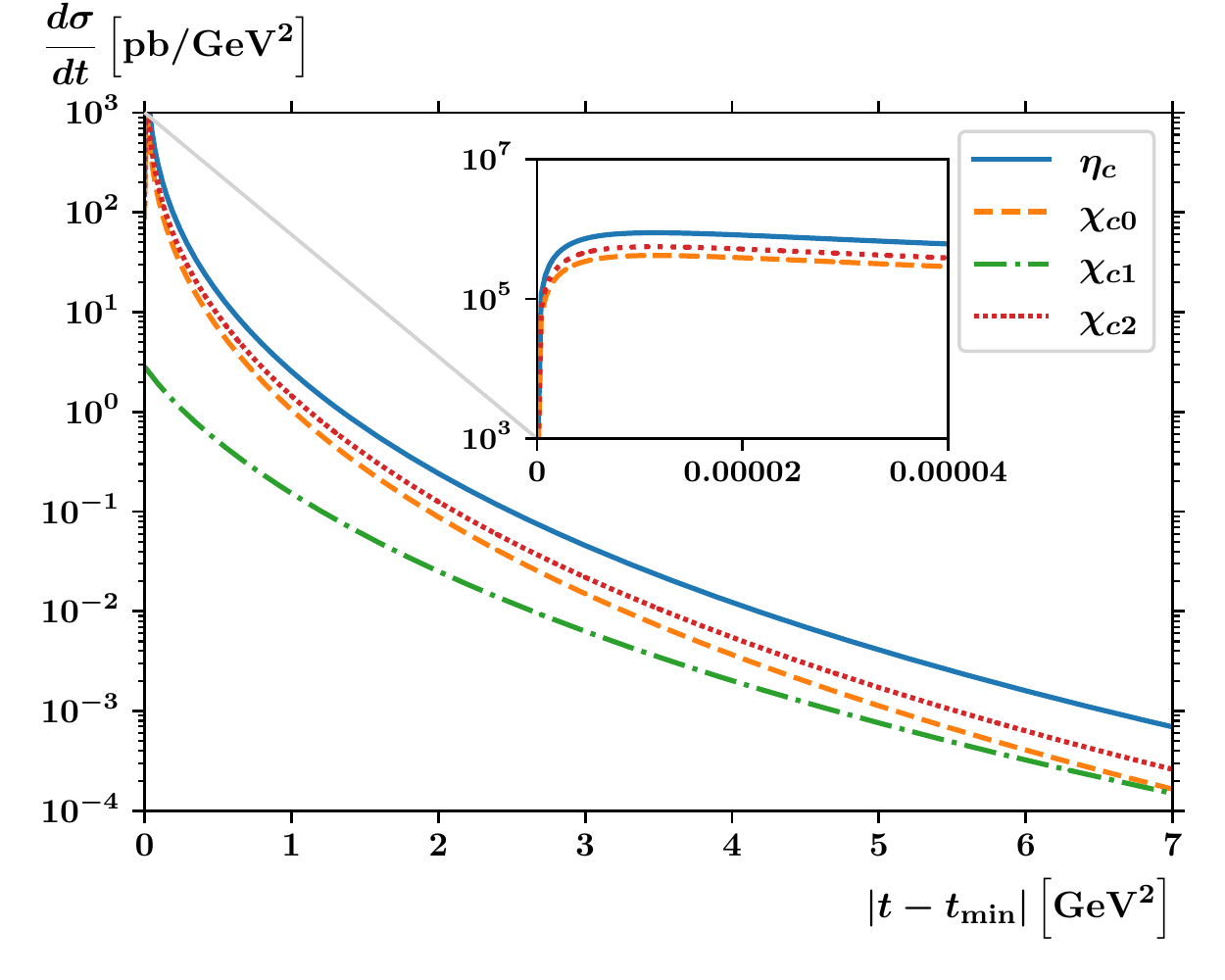}\label{fig:cross_c_photon_W50}}

  \caption{The differential production rates for $C$-even charmonia at different center-of-mass energy, (a),  (b), (c). }
  \label{fig:cross_c_photon}
\end{figure}
\begin{figure}[htb]
  \subfloat[$\sqrt{s}=15\:\text{GeV}$]{\includegraphics[scale=0.6]{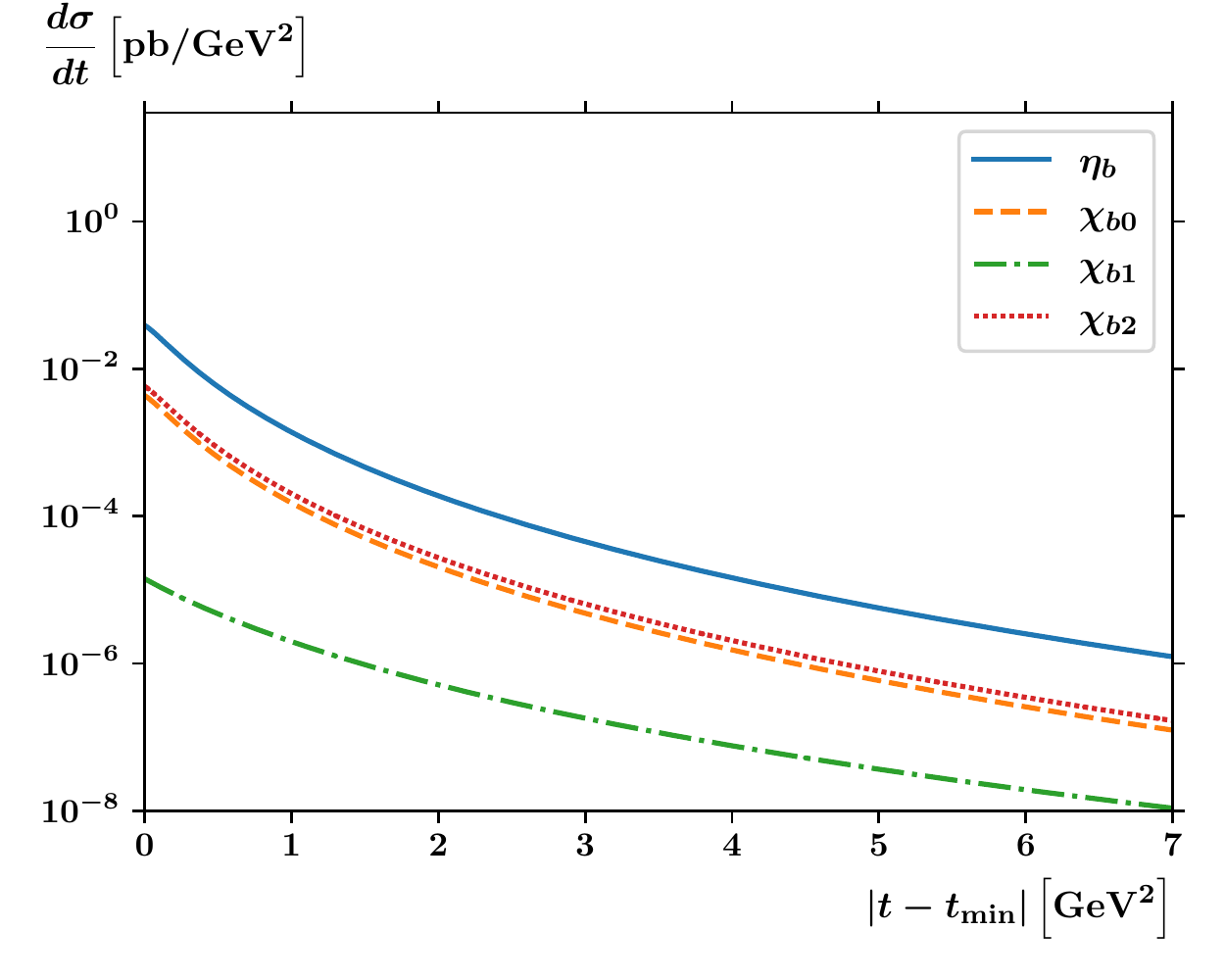}}
  \subfloat[$\sqrt{s}=50\:\text{GeV}$]{\includegraphics[scale=0.6]{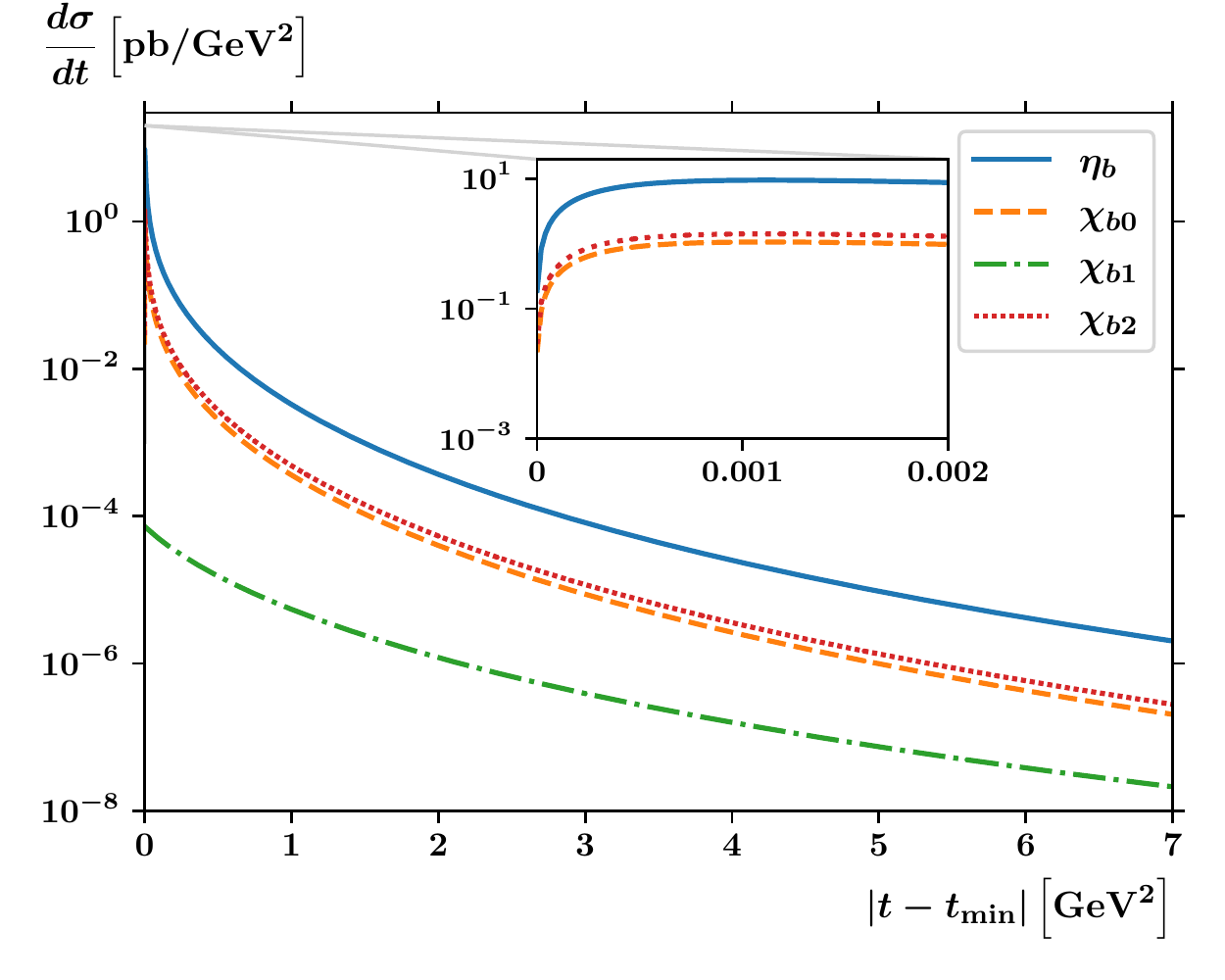}\label{fig:cross_b_photon_W50}}

  \caption{The differential production rates for $C$-even bottomonia at different $\sqrt{s}$, (a), (b). }
  \label{fig:cross_b_photon}
\end{figure}

In FIG. \ref{fig:cross_c_photon} and FIG.
\ref{fig:cross_b_photon} we show the profiles differential photoproduction rates of $C$-even charmonia and bottomonia
at several benchmark points of center-of-mass energy.
Typically, we choose $\sqrt{s}=4.3\:\text{GeV}$ (near charmonium threshold), $15\:\text{GeV}$
(near bottomium threshold), and $50\:\text{GeV}$ (high energy limit).
\begin{figure}[htb]
  \includegraphics[scale=0.6]{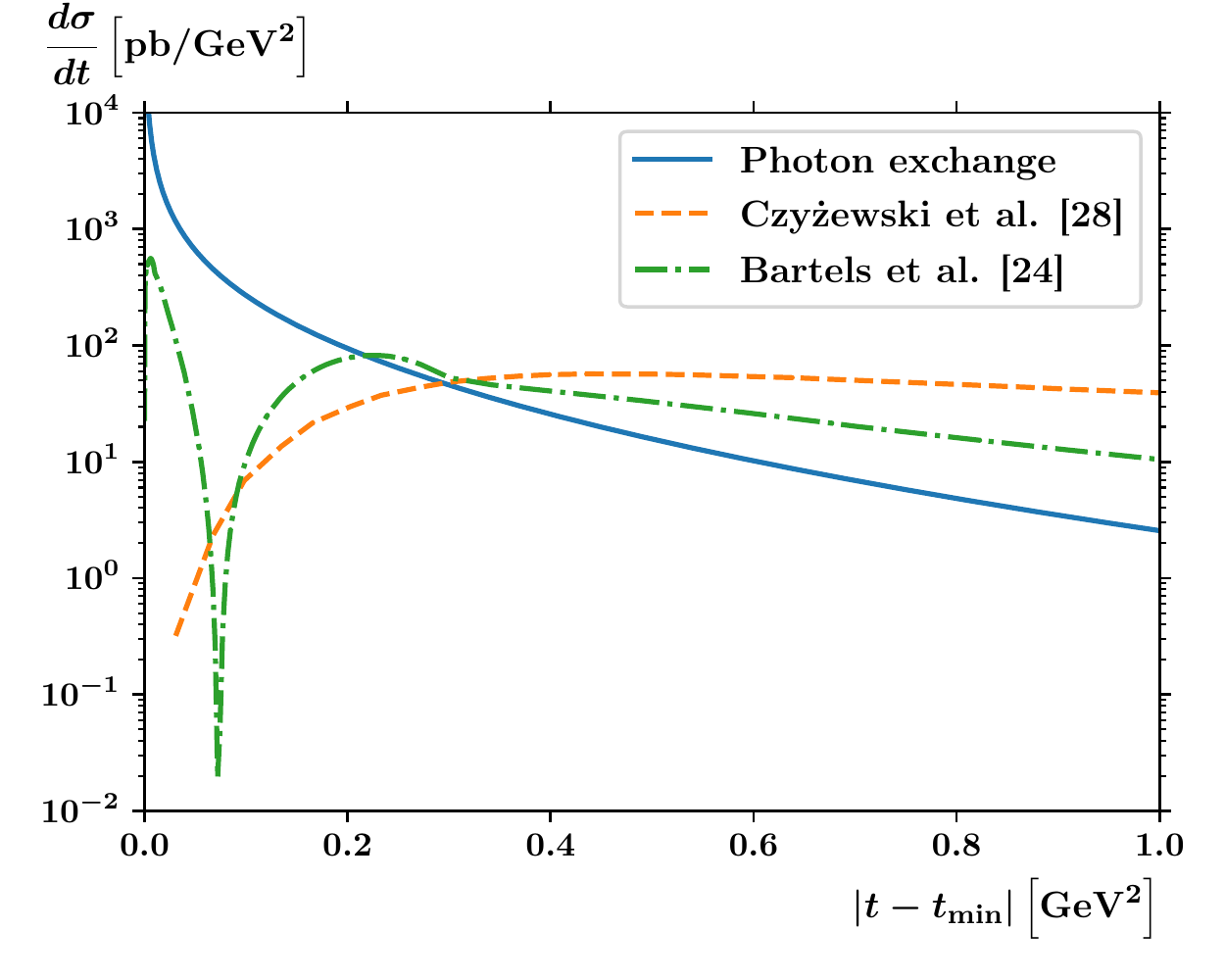}
  \caption{The differential photoproduction rates of $\eta_c$ at $\sqrt{s}=300\:\text{GeV}$ via the one-photon exchange mechanism, 
  compared with the prediction from two Odderon exchange models~\cite{Czyzewski:1996bv} and \cite{Bartels:2001hw}. }
  \label{fig:compare_Odderon}
\end{figure}

Remarkably, in FIG.~\ref{fig:compare_Odderon} we also make a detailed comparative study between the one-photon exchange prediction the Odderon
model predictions~\cite{Czyzewski:1996bv, Bartels:2001hw} for $\eta_c$ photoproduction. 
We take the high energy limit $\sqrt{s}=300\:\text{GeV}$.
Curiously, one finds that in the low-$|t|$ region, the one-photon exchange contribution
is comparable with in magnitude, or even larger than, the Odderon exchange contributions. 
This implies that it is mandatory to include the one-photon exchange contribution in confrontation from the 
future experimental measurements, even though the ultimate goal is to probe the Odderon contribution
in an unambiguous way.
\begin{figure}[htb]
  \subfloat[charmonia]{\includegraphics[scale=0.6]{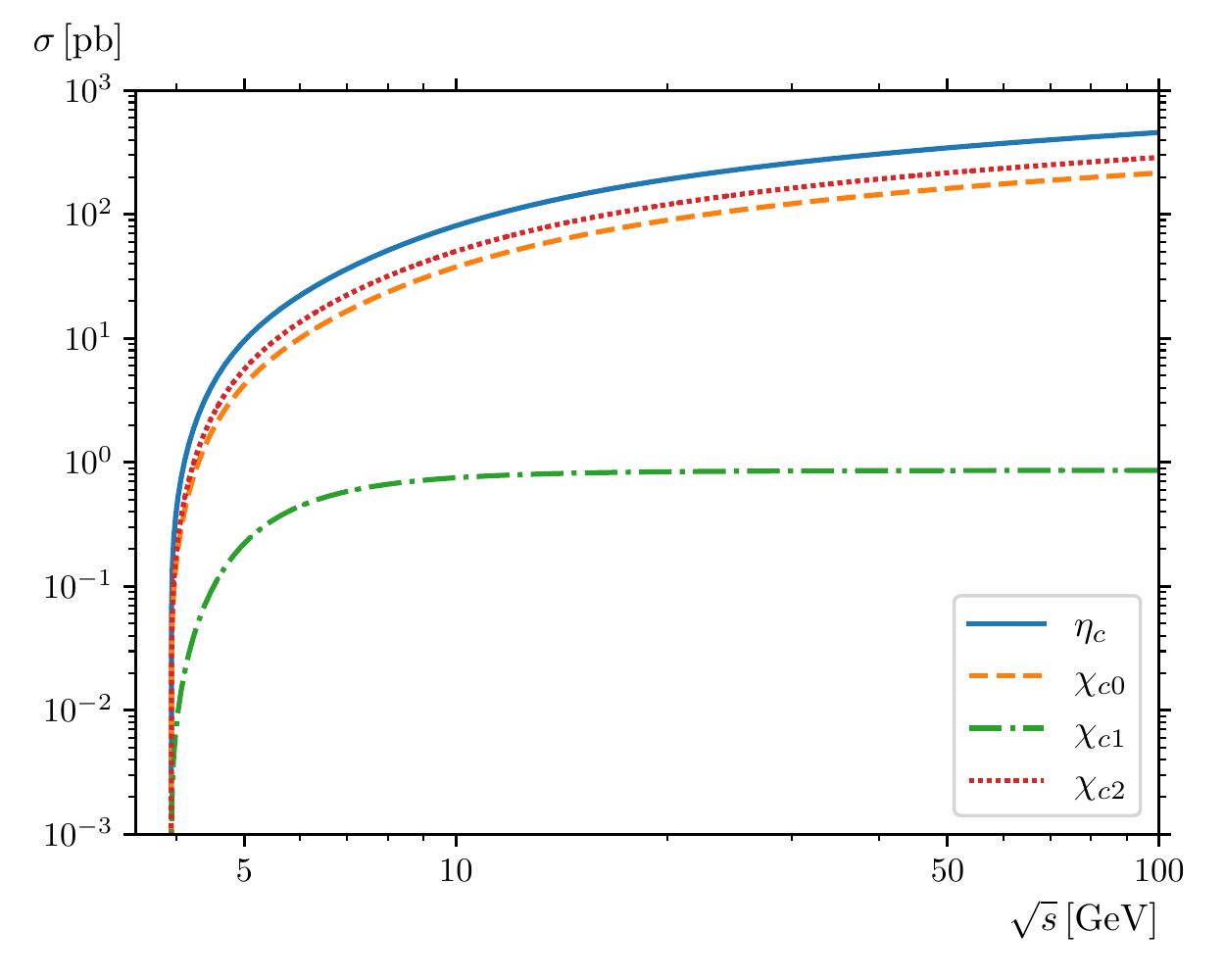}\label{fig:cross_c_photon_total}}
  \subfloat[bottomonia]{\includegraphics[scale=0.6]{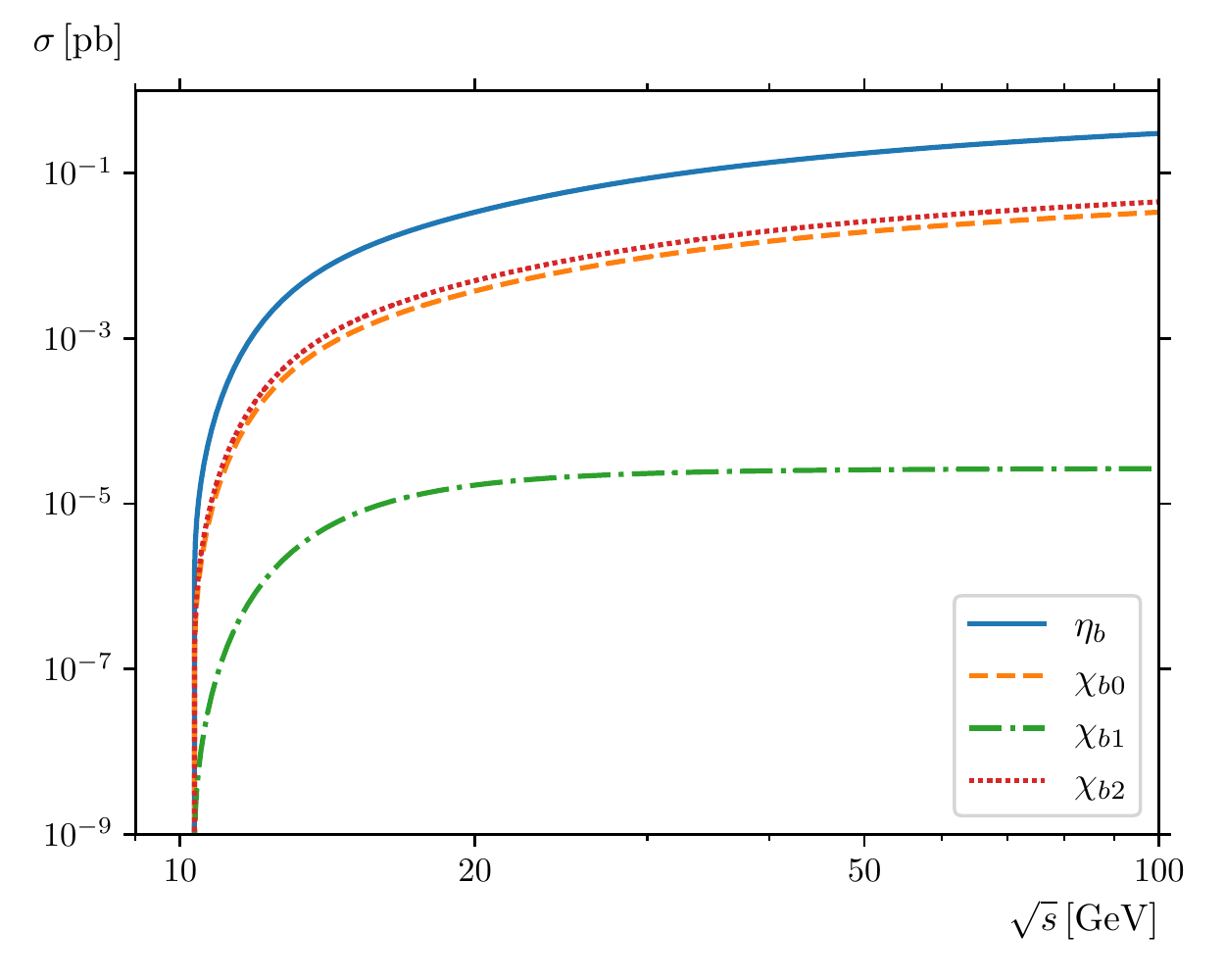}\label{fig:cross_b_photon_total}}

  \caption{The integrated photoproduction rates for $C$-even charmonia $(a)$ and bottomonia $(b)$. }
  \label{fig:cross_photon_total}
\end{figure}

The integrated photoproduction cross sections as a function of $\sqrt{s}$ are displayed in FIG.~\ref{fig:cross_photon_total}. 
One can see that the total cross sections for $\eta_c$ and $\chi_{c0,2}$ photoproduction are of order $10^2\:\text{pb}$
in large $\sqrt{s}$, whereas the cross sections for $\eta_b$ and $\chi_{b0,2}$ are much smaller, roughly of order $10^{-2}\sim10^{-1}\:\text{pb}$. 
Compared with the $\chi_{c(b)0,2}$ channels, the cross sections for $\chi_{c(b)1}$ photoproduction 
are much suppressed at high energy. This phenomenon can be explained as follows. 
Clearly the main contribution to the total photoproduction cross section comes from the low-$|t|$ region, where the
exchanged photon becomes quasi-real. In such a limit, one can utilize the
equivalent photon approximation (EPA). The metric tensor in the numerator of the photon propagator
(FIG. \ref{fig:diagrams_photon}) can be reexpressed in terms of four photon polarization vectors
%-------------------------------
\beq
%-------------------------------
 -g^{\mu\nu} = -\varepsilon^\mu_+(q) \varepsilon^{\nu*}_-(q) - \varepsilon^\mu_-(q) \varepsilon^{\nu*}_+(q) +
  \sum_i \varepsilon^\mu_{Ti}(q) \varepsilon^{\nu*}_{Ti}(q).
%-------------------------------
\eeq
%-------------------------------
$\varepsilon^\mu_{Ti}(q)$ are transverse polarization vectors, defined in a frame where
$\varepsilon^\mu_+(q) = (1,0,0,1)^T \propto q^\mu$ and $\varepsilon^\mu_-(q) = (1,0,0,-1)^T$. The
contraction between $\varepsilon^{\mu(*)}_+(q)$ with $\langle H\vert (J_\text{EM})_{\mu}\vert\gamma\rangle$ or
$\langle p'\vert (J_\text{EM})_{\mu}\vert p\rangle$ in \eqref{eq:iM} yields vanishing contributions by the Ward identity. 
The contraction between $\varepsilon^\mu_{Ti}(q)$ with $\langle \chi_{c(b)1}\vert (J_\text{EM})_{\mu}\vert\gamma\rangle$ also gives zero in
the $t\rightarrow 0$ limit, since $\chi_{c1}\to\gamma\gamma$ is forbidden by the Landau-Yang theorem. Therefore 
the amplitude for $\chi_{c1}$ photoproduction is severely suppressed in the small $|t|$ limit. 
A direct calculation shows this suppression gives rise to an overall
$t$ factor that cancel the denominator in \eqref{eq:iM}. The explicit scaling behavior with $t$ for quarkonium photoproduction
can also be seen from the denominators of differential cross sections in \eqref{eq:dsigma_asym}.

\subsection{Equivalent photon approximation}

In order to estimate the quarkonium photoproduction rates in realistic 
$ep$ collision experiments such as \texttt{EIC} and \texttt{EicC}, one often invokes the equivalent photon approximation~\cite{Budnev:1974de}
to assume the photon emitted off the incident electron carrying small virtuality $Q^2$.
The exclusive quarkonium production rate in $ep$ collision can be factorized as \cite{Xie:2021seh}
%-------------------------------
\beq
%-------------------------------
  \sigma\left(ep\rightarrow eHp\right)\approx\int_{k_\text{min}}^{E_e} dk\int_{Q_\text{min}}^{Q_\text{max}} dQ^{2}\frac{d^{2}N_{\gamma}}{dkdQ^{2}}\sigma\left(\gamma p\rightarrow Hp\right),
%-------------------------------
  \label{eq:EPA}
%-------------------------------
\eeq
%-------------------------------
with the photon flux defined as
\beq
  \frac{d^{2}N_{\gamma}}{dkdQ^{2}}=\frac{\alpha}{\pi kQ^{2}}\left[1-\frac{k}{E_{e}}+\frac{k^{2}}{2E_{e}^{2}}-\left(1-\frac{k}{E_{e}}\right)\frac{Q_{\text{min}}^{2}}{Q^{2}}\right],
\eeq
where $k$ and $E_{e}$ are the photon and electron energies in the target rest frame. $k_\text{min}=E_e M_H\left(M_H+2M_p\right)/\left(s-M_p^2\right)$,
$Q_{\text{min}}^{2}=m_{e}^{2}k^{2}/\left(E_{e}\left(E_{e}-k\right)\right)$.  
We choose $Q_\text{max}^2=0.01\:\text{GeV}^2$ to be the criterion for identifying a photoproduction event~\footnote{According to 
\cite{AbdulKhalek:2021gbh}, choosing $Q_\text{max}^2$ to be 0.01 or 0.1 does not affect
the order of magnitude of the estimated cross section.}.
We take the $\chi_{c(b)0}$ photoproduction rates as a reference basis for the estimation. 
The cross sections for $\chi_{c(b)2}$ are roughly equal to that for $\chi_{c(b)0}$,  while the cross section for 
$\eta_c$ is roughly twice that of $\chi_{c0}$, but the cross section of $\eta_b$ is roughly 10 times larger than that
of $\chi_{b0}$.

After carrying out the numerical integration in \eqref{eq:EPA} at $\sqrt{s}=50$ GeV, the typical center-of-mass 
energy at \texttt{EIC}, the photoproduction rates in $ep$ collision
can reach $7.5\:\text{pb}$ for $\chi_{c0}$ and $0.35\:\text{fb}$
for $\chi_{b0}$. Concerning the projected high integrated luminosity of
$50\:\text{fb}^{-1}/\text{year}$ at \texttt{EicC}~\cite{Anderle:2021wcy} and $1.5\:\text{fb}^{-1}/\text{month}$ at \texttt{EIC}~\cite{AbdulKhalek:2021gbh}, 
these $C$-even quarkonia photoproduction processes have bright prosect to be observed in the future \texttt{EicC} and  \texttt{EIC} experiments.

For comparison, we also estimate photoproduction rate $\eta_c$ in the near threshold limit. We assume $\sqrt{s}=4.6\:\text{GeV}$, 
which corresponds to $E_\gamma\approx11\:\text{GeV}$ in the proton rest frame, which lies within the 
the energy range of the \texttt{GlueX} experiment($\sqrt{s}<6$ GeV, with $E_\gamma<15$ GeV~\cite{Ali:2019lzf}).
The $\eta_c$ production rate is estimated to be about $5\:\text{pb}$. Since the integrated luminosity
of the \texttt{GlueX} experiment is only about $68\;\text{pb}^{-1}$~\cite{Ali:2019lzf}, and the reconstruction
efficiency for $\eta_c$ and $\chi_{cJ}$ is much lower than $J/\psi$, it is difficult to observe these photoproduction processes of $C$-even
charmonia in the current JLab facilities.

\section{Conclusion}
\label{sec:conclusion}

In the present paper, we analyze the photoproduction of $C$-even quarkonia via the one-photon exchange mechanism. 
The main result is that even in the Regge limit, the photoproduction rates for most channels through this mechanism are comparable or even greater than 
the predictions based on Odderon exchange models\cite{Czyzewski:1996bv,Bartels:2001hw}. They have bright prospects to be observed in
future \texttt{EIC} and \texttt{EicC} experiments. 
As an interesting exception, the photoproduction rates for $\chi_{c/b,1}$ are much
suppressed relative to other channels in the Regge limit, as a consequence of Landau-Yang theorem. Due to the lacking of one-photon exchange contribution, 
perhaps the future observation of the $\chi_{c/b,1}$ photoproduction events may be viewed as a strong evidence for the Odderon exchange model.

\appendix

\section{One-gravition exchange for $J/\psi$ photoproduction}
\label{sec:graviton}

Understanding the origin of proton mass from QCD has been a fascinating topic for
years. The QCD trace anomaly is believed to make important contribution to the proton mass. 
It has been proposed that $J/\psi$ photoproduction near threshold could be a golden channel to infer the QCD trace anomaly contribution
to the proton mass~\cite{Kharzeev:1998bz}. However, it has also been criticized that some phenomenological assumptions such as vector meson dominance
have been adopted in \cite{Kharzeev:1998bz}.

A closely related topics is the gravitational form factor (GFF) of the proton, from which one can deduce the mass and stress/pressure distributions of
the proton. It is rather difficult to extract the proton GFF directly from the scattering experiment. So far the best knowledge about the proton GFF
is from lattice QCD simulation~\cite{Shanahan:2018pib, Azizi:2019ytx}. 

\begin{figure}[h]
  \includegraphics{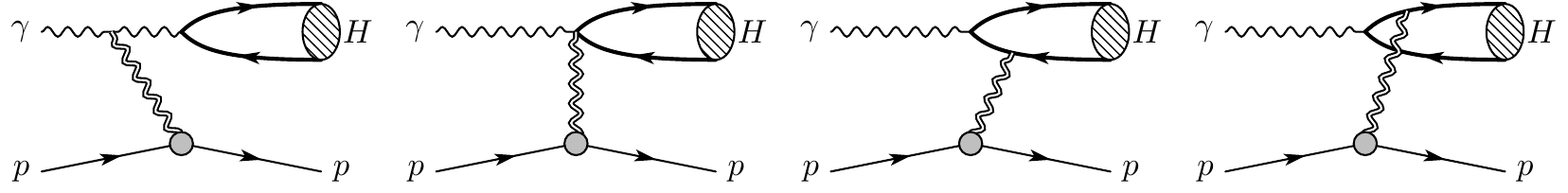}
  \caption{Feynman diagrams in the LO gravitational contribution. }
  \label{fig:diagrams_graviton}
\end{figure}

In this appendix, analogous to $\gamma p\rightarrow \eta_c p$  process via the one-photon exchange, 
we calculate the photoproduction process $\gamma p\rightarrow J/\psi p$
through one-graviton exchange. The corresponding LO Feynman diagrams are shown in 
shown in FIG.~\ref{fig:diagrams_graviton}. It is certainly hopefulness to observe this graviton-induced process
in any terrestrial collision experiments, we include the corresponding results mainly for amusement. Note that unlike
the photon which is $C$-odd, the graviton is $C$-even, therefore the final quarkonium has to be the $C$-odd state
such as $J/\psi$.

The corresponding photoproduction amplitude can be written in a product of matrix elements
of the symmetric (Belinfante) energy-momentum tensor $\Theta_{\mu \nu}$:
%-------------------------------
\beq
%-------------------------------
  i \mathcal{M}= \langle J/\psi(P) \vert \Theta_{\mu \nu} \vert \gamma(k)\rangle \frac{-i \mathcal{P}^{\mu \nu \rho \sigma}}{q^2} \langle p(P_2) \vert \Theta_{\rho \sigma} \vert p(P_1)\rangle,
%-------------------------------
\eeq
%-------------------------------
with $\mathcal{P}^{\mu \nu \rho \sigma}=\frac{1}{2}\left(g^{\mu\rho}g^{\nu\sigma}+g^{\mu\sigma}g^{\nu\rho}-g^{\mu\nu}g^{\rho\sigma}\right)$.  
The corresponding matrix element of $\Theta_{\mu \nu}$ for proton is nothing but the proton GFF, which is often parameterized as 
%-------------------------------
\begin{align}
%-------------------------------
  & \langle p(P_2) \vert \Theta_{\mu \nu} \vert p(P_1)\rangle =\kappa \bar{u}(P_2)[\,A(t)\, \frac{P_{\mu}\,P_{\nu}}{M_p}\,+B(t)\, \frac{i\, P_{\lbrace \mu} \sigma_{\nu \rbrace \rho}\Delta^{\rho}}{2M_p}\,+D(t)\frac{\Delta_{\mu}\Delta_{\nu}-\eta_{\mu \nu}\Delta^{2}}{4M_p}]u(P_1),\\
  & \quad P=\frac{1}{2}(P_1+P_2), \quad \Delta=P_2-P_1, \quad t=\Delta^{2},
\end{align}
%-------------------------------
where the Lorentz scalars $A$, $B$, $D$ are functions of the momentum transfer.

The $\gamma$-to-$J/\psi$ gravitational transition form factor can also be computed in NRQCD factorization. At the lowest order in $\alpha_s$ and $v$, 
it reads
%-------------------------------
\begin{flalign}
%-------------------------------
 & \Theta^{\mu \nu}= \bar{\psi}\gamma^{(\mu}i\overleftrightarrow{D}^{\nu)}\psi+\frac{1}{4}g^{\mu \nu}F^2-F^{\mu \rho}F^{\nu}_{\rho},\\
 & \langle J/\psi(P) \vert \Theta^{\mu \nu} \vert \gamma(k)\rangle=\frac{1 }{6 m_{c}^{3/2} (k\cdot P  )}\sqrt{\dfrac{N_c}{2\pi}}R_S(0)\varepsilon^{*}_{ \sigma}\left(P\right)\,\varepsilon_{\rho}\left(k\right)\nonumber\\
    & e \kappa \left(g^{\mu  \nu } g^{\rho  \sigma } (k \cdot P  )^2-g^{\mu  \rho } g^{\nu  \sigma } (k \cdot P )^2-k^{\sigma } P^{\rho } g^{\mu  \nu } (k\cdot P  )+k^{\sigma } P^{\nu } g^{\mu  \rho }(k\cdot P  )\right.\nonumber\\
    & \left.+k^{\sigma } P^{\mu } g^{\nu  \rho } (k\cdot P  )+g^{\mu  \sigma } \left(2 m_{c}^2-k\cdot P  \right) \left(g^{\nu  \rho } (k\cdot P  )-k^{\nu } P^{\rho }\right)-k^{\nu } P^{\mu } g^{\rho  \sigma } (k\cdot P  )\right.\nonumber\\
    & -2 m_{c}^2 k^{\nu } k^{\sigma } g^{\mu  \rho }+k^{\mu } \left(-2 m_{c}^2 k^{\sigma } g^{\nu  \rho }+4 m_{c}^2 k^{\nu } g^{\rho  \sigma }+P^{\rho } g^{\nu  \sigma } \left(k\cdot P  -2 m_{c}^2\right)\right.\nonumber\\
    &\left.\left. -P^{\nu } g^{\rho  \sigma } (k\cdot P )\right)+2 m_{c}^2 g^{\mu  \rho } g^{\nu  \sigma } (k\cdot P  )\right),\\
       & k\cdot P=2m_c^2-\frac{t}{2},\nonumber
%-------------------------------
\end{flalign}
%-------------------------------
where $k$ is the incoming momentum of the photon and $P$ is the momentum of the outgoing $J/\psi$. 
One can readily verify that the above expressions obey Ward identity and the conservation of $\Theta_{\mu \nu}$.

After some straightforward calculation, the differential photoproduction rate of $J/\psi$ via graviton exchange reads
%-------------------------------
\begin{align}
%-------------------------------
  & \frac{d\sigma}{dt}_\text{graviton} =\frac{e^2 \kappa^4 N_c \left|R_S(0)\right|^2}{288 \pi M_{p}^2 m_{c}^3 \left(t-4 m_{c}^2\right)^2}\,\left(A^2(t) \left(4 M_{p}^2-t\right) \Big(M_{p}^4-2 M_{p}^2 s+t \left(s-m_{c}^2\right)\right.\nonumber\\
  &+\left(s-2  m_{c}^2\right)^2\Big)^2 +2 A(t) \left(D(t) m_{c}^4 \left(4 M_{p}^2-t\right)\left(-2 M_{p}^2-4 m_{c}^2+2 s+t\right)^2\right.\nonumber\\
  & +2
  B(t) t \left.\left(M_{p}^4-2 M_{p}^2 s+t \left(s-m_{c}^2\right)+\left(s-2 m_{c}^2\right)^2\right)^2\right)\nonumber\\
  & +D^2(t) m_{c}^4 \left(4 M_{p}^2-t\right) \left(t-4 m_{c}^2\right)^2+4 D(t) B(t) m_{c}^4 t \left(-2 M_{p}^2-4 m_{c}^2+2 s+t\right)^2\nonumber\\
  & -4 B^2(t) \left(M_{p}^8 t+2 M_{p}^6 \left(8 m_{c}^4+m_{c}^2 t-t (2 s+t)\right)+2 M_{p}^4 \left(16 m_{c}^6-8 m_{c}^4 (2 s+t)\right.\right.\nonumber\\
  & \left.+m_{c}^2 t (t-6 s)+3 s t (s+t)\right)+M_{p}^2 \left(64 m_{c}^8-16 m_{c}^6 (4 s+t)+4 m_{c}^4 \left(4 s^2-2 s t-t^2\right)\right.\nonumber\\
  & +m_{c}^2 t\left.\left.\left.\left(18 s^2+14 s t+t^2\right)-2 s t (s+t) (2 s+t)\right)+t \left(t \left(s-m_{c}^2\right)+\left(s-2 m_{c}^2\right)^2\right)^2\right)\right)
%-------------------------------
\end{align}
%-------------------------------
where $\kappa^{2}=32\pi G_N=6.75\times10^{-37}\:\text{GeV}^{-2}$.
Adopting the lattice QCD estimates of the proton GFF~\cite{Shanahan:2018pib, Azizi:2019ytx}, we find that the integrated cross section
for $\gamma p\to J/\psi p$ is about 66 orders of magnitude smaller than that $\gamma p\to \eta_c p$ via one-photon exchange (See FIG.9)!

\begin{figure}[h]
  \includegraphics[scale=0.5]{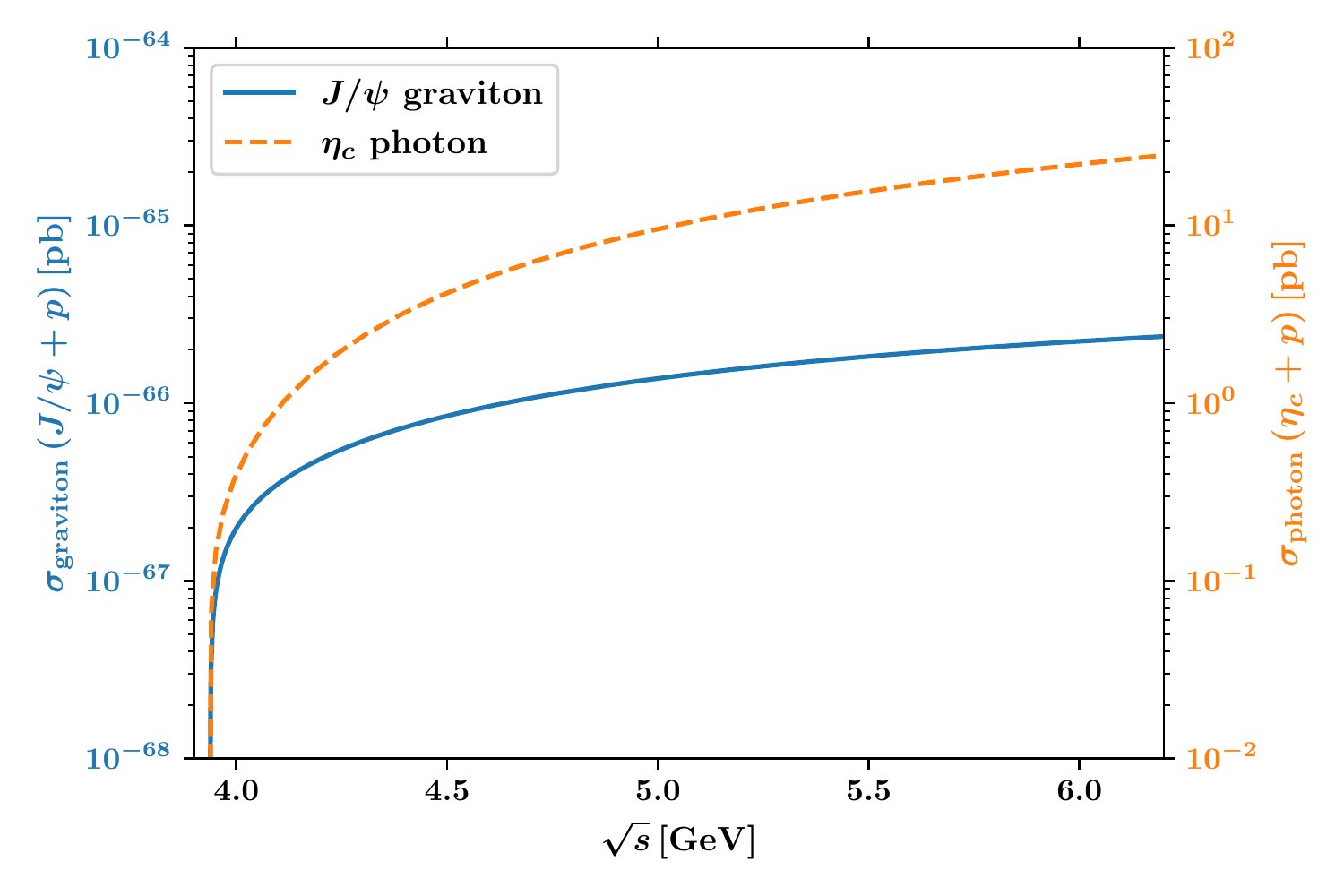}
  \caption{Comparison of the photoproduction rates between photon induced $\eta_c$ production and graviton-induced $J/\psi$ production.}
  \label{fig:graviton}
\end{figure}

%-------------------------
\begin{acknowledgments}
%-------------------------
%-------------------------
This work is supported in part by the National Natural Science Foundation of China under Grants No.~11925506, 11875263, No.~11621131001 (CRC110 by DFG and NSFC).
%-------------------------
\end{acknowledgments}
%-------------------------

\end{document}